\documentclass[pra,final,showpacs,superscriptaddress,aps]{revtex4-2}
\pdfoutput=1
\usepackage{lipsum}
\usepackage{amssymb}
\usepackage{amsmath}
\usepackage{color}
\usepackage{multirow}
\usepackage{afterpage}
\usepackage{dsfont}
\usepackage{braket}
\usepackage{ntheorem}
\usepackage [misc]{ifsym}

\begin{document}
\title{Measurement-induced nonbilocal correlation based on Wigner-Yanase skew information}

\author{Jianhui Wang}
\email{12020117009@mail.ynu.edu.cn}
\author{Yajie Wang}
\author{Qing Chen}
\email{chenqing@ynu.edu.cn}
\affiliation{School of Physics and Astronomy,Yunnan University, Kunming 650000, China}
\affiliation{Key Laboratory of Quantum Information of Yunnan Province, Kunming 650500, China}

\begin{abstract}
Measurement-induced nonlocality (MIN) was proposed for measure the maximum global effect caused by locally invariant measurements. Similarly, the Measurement-induced nonbilocal correlation is a generalization of MIN can be used to measure the maximal global influence caused by the local measurement in the bilocal scenario. In this paper, we propose a new nonbilocal correlation measure based on the Wigner-Yanase skew information. The relationship between the MIN based on Wigner-Yanase skew information and our measure is discussed. We present an analytical expression of our measure for pure input states and also provide upper bounds for input general mixed states. 
\end{abstract}

\maketitle

\section{Introduction}
One of the most important problem in quantum physics is to characterize the correlations between separated observers, and the original disscusion about correlations in the quantum physics can be traced back to EPR paradox\cite{Einstein1935}. The study of nonlocal correlation has fundamental implication for quantum information and other related fields.  The latest progress in this field\cite{Branciard2010, Branciard2012, Tavakoli2014, Renou2019, Munshi2021, Pozas2022} is mainly focused on  rely entanglement to refute local hidden variable theory, based on the famous Bell theorem\cite{Bell1964, Brunner2014}, to show the nonlocal correlations. However, it is well known that entanglement and nonlocality are different resources\cite{Brunner2005} and entanglement-separability verification is not the only way to explore the nonlocal correlation. An alternative trial way is to find a kind of metric to measure quantum correlation. The quantum discord as such a method was initially introduced by Olliver and Zurek\cite{Ollivier2002} and by Henderson and Vedral\cite{Henderson2001}, and deveopled in the last decade\cite{Luo2010, Daki2011, Luo2012, Piani2012, Chang2013, Bera2018}. The measurement-induced nonlocality(MIN)\cite{Luo2011} as another useful measure is different from, in some extend dual to, the geometric measure of quantum discord\cite{Luo2010} and can be used to capture the genuine nonlocal effect of measurements on the state. The original proposed MIN is based on the Hilbert-Schmidt norm and now is extend to other measure\cite{Xi2011, Hu2015, Li2016, Muth2017, Muth2019,VS2021}.

The quantum nonlocality not only exist in the two-observer situation but also be found in multi-observer and multi-source scenarios. In particular, the simplest and nontrivial example is the entanglement-swapping experiment that is the foundation of more complicated quantum network and has be wildly studied\cite{Branciard2010, Branciard2012, Thomas2001, Short2005, Gisin2017, Andreoli2017, Carvacho2017, Saunder2017, Sun2018}. There are some distinct features in this scenario, such as exist states that cannot vioalte the Clauser-Horne-Shimony-Holt (CHSH) inequality\cite{Clauser1970} but can violate the bilocal inequality\cite{Branciard2010, Branciard2012, Tavakoli2021}, ie., there exist nonbilocal correlations. To capture the nonbilocal correlations, there are some extensions of MIN applied to the bilocal scenario are proposed\cite{Zhang2021, Zhang2021G, Bhuvaneswari2022, Muthuganesan2022}.

In this paper, we propose a kind of measurement-induced nonbilocality measure based on the Wigner-Yanase skew information. This paper is organized as follows. In Section 2, we brefily introduce  two nonlocality measures, MIN and MIN based on Wigner-Yanase skew information(MINS)\cite{Li2016}. In Section 3, we define a new measure that is the measurement-induced nonbilocality based on Wigner-Yanase skew information (MINBS) that can be viewd as a kind of generalization of MINS, and we give some basic properties of it. We also obatained the analytical solutions of the pure states and the tight upper bound of the mixed states. In Section 4, we calculate some examples to show this nonbilocality measure. At the end, we summarize our results in Section 5.

\section{Measurement-Induced Nonlocality}

In contrast with the Bell nonlocality, MIN is in some kind of more general correlation. For capture all the effects that can be induced by local measurements, MIN is defined as

\begin{equation}
N(\rho_{ab)\equiv\rm{\mathop{max}\limits_{\Pi^a}}\Vert \rho_{ab}-\Pi^a(\rho_{ab})\Vert^2},
\end{equation}

where the maximum is taken over all the von Neumann measurements $\Pi^a=\{\Pi^a_k\}$ which do not disturb $\rho^a$ locally, that is $\sum_k\Pi^a_k\rho^a\Pi^a_k=\rho^a$, and $\Vert \cdot\Vert^2$ in the original proposal[19] is taken as the Hilbert-Schmidt norm $\Vert X\Vert^2=trX^\dagger X$,the $\Pi^a(\rho_{ab})$ denots the post-measurement state, this quantity is an indicator of the global effect caused by locally invariant measurements.

In particular, as proposed by Piani for the geometric discord[16], the MIN also may change rather arbitrarily through some trivial and uncorrelated action of unmeasured party b. To see this regarding the state $\rho^{a:bc}=\rho^{a:b}\otimes\rho^c$ as a bipartite state; then 
$$
N(\rho^{a:bc})=N(\rho^{a:b})tr(\rho^c)^2,
$$

which, as long as $\rho^c$ is a mixed state, differs from the intuitive requirement $N(\rho^{a:bc})=N(\rho^{a:b})$. To avert this issue, we can take measure that have contractivity such as the Wigner-Yanase skew information. The MIN based on Wigner-Yanase skew information is defined as[22]

\begin{equation}\label{2}
	\mathcal{N}_s(\rho^{ab}) = \mathop{max}\limits_{\{\Pi^k_a\}}\sum\limits^m_{k=1}\mathcal{I}(\rho^{ab},\Pi^k_a\otimes\mathbb{I}_n),
\end{equation}

where $\mathcal{I}(\rho,K)$ is the skew information introduced by Wigner and Yanase\cite{Wigner1963} and $\mathcal{I}(\rho,K)=-\frac{1}{2}tr[\rho^{\frac{1}{2}},K]^2$ where $K$ is a Hermitian observable and [,] denots the commutator.Tha max is taken over all the von Neumann measurements which do not disturb $\rho^a$ locally.

\section{Measurement-induced Nonbilocality based on Wigner-Yanase skew information}
In this section, we introduce our nonbilocality measure based on Wigner-Yanase skew information.

For any arbitrary finite dimensional system $H= H_A\otimes H_B\otimes H_C\otimes H_D$ with $dim\ H_A=m$, $dim\ H_B=n$, $dim\ H_C=u$ and $dim\ H_D=v$.

Define the Measurement-induced nonbilocality based on Wigner-Yanase skew information as 

\begin{equation}\label{3}
	\mathcal{N}^b_s(\rho_{AB}\otimes \rho_{CD}) = \mathop{max}\limits_{\{\Pi^{BC}_{st}\}}\sum\limits^n_{s=1}\sum\limits^u_{t=1}\mathcal{I}(\rho_{AB}\otimes \rho_{CD},\mathbb{I}_m\otimes\Pi^{BC}_{st}\otimes\mathbb{I}_v).
\end{equation}

Noting that this formula can be reduced to 

\begin{align}
	\mathcal{N}^b_s(\rho_{AB}\otimes \rho_{CD})=1-&\mathop{min}\limits_{\{\Pi^{BC}_{st}\}}\sum\limits^n_{s=1}\sum\limits^u_{t=1}tr(\sqrt{\rho_{AB}\otimes \rho_{CD}}(\mathbb{I}_m\otimes\Pi^{BC}_{st}\otimes\mathbb{I}_v) \sqrt{\rho_{AB}\otimes \rho_{CD}}(\mathbb{I}_m\otimes\Pi^{BC}_{st}\otimes\mathbb{I}_v),
\end{align}

where the $min$ is taken over all the von Neumann measurements which do not disturb $\rho^{bc}$ locally.

There we list some basic properties of this measure.
\begin{itemize}
	\item [(i)]
	$\mathcal{N}^b_s(\rho_{AB}\otimes \rho_{CD})=0$  for all product states $\rho_{AB}=\rho_A\otimes \rho_B$ and $\rho_{CD}=\rho_C\otimes \rho_D$. But this measure is strictly positive for $\rho_{AB}, \rho_{CD}$ where at least one of the states is entanglement.
	
	In particular, we have
	
	\item [(ii))]
	$\mathcal{N}^b_s(\rho_{AB}\otimes \rho_{CD})$=0 for any quantum-classical state\cite{Luo2008} $\rho_{AB}=\sum_k\rho^A_k\otimes p_k\ket{k_B}\bra{k_B}$ and classical-quantum state $\rho_{CD}=\sum_f q_f\ket{f_C}\bra{f_C}\otimes\rho^D_f$ whose marginal states $\rho_B$ and $\rho_A$ are both nondegenerate.
	
	\item [(iii)]
	If $\rho_B, \rho_C$ have the spectral decomposition $\rho_B=\sum_i \lambda_i\ket{i_B}\bra{i_B}, \rho_C=\sum_j \mu_j \ket{j_C}\bra{j_C}$, respectively, and at least one of the states $\rho_B$ and $\rho_C$ is nondegenerate, then the von Neumann measurements that do not disturb $\rho_{BC}$ must have the form $\Pi^{BC}=\{\Pi^B_i\otimes \Pi^C_j\}$, ie., they are not the joint quantum measurements.
	
	\item [(iv)]
	$\mathcal{N}^b_s(\rho_{AB}\otimes \rho_{CD})$ is locally invariant in the sense that $\mathcal{N}^b_s((U_A\otimes U_B\otimes U_C\otimes U_D)\rho_{AB}\otimes \rho_{CD}(U_A\otimes U_B\otimes U_C\otimes U_D)^\dagger)=\mathcal{N}^b_s(\rho_{AB}\otimes \rho_{CD})$ for any unitary operators $U_A,U_B,U_C,U_D$ acting on $H_A,H_B,H_C,H_D$, respectively.
	
	\item [(v)]
	If $\rho_B$ and $\rho_C$ are both nondegenerate, then
	\begin{align}
		\mathcal{N}^b_s(\rho_{AB}\otimes \rho_{CD})=1-&\sum\limits_{st}tr(\sqrt{\rho_{AB}\otimes \rho_{CD}}(\mathbb{I}_m\otimes\Pi^{BC}_{st}\otimes\mathbb{I}_v)\sqrt{\rho_{AB}\otimes \rho_{CD}}(\mathbb{I}_m\otimes\Pi^{BC}_{st}\otimes\mathbb{I}_v).\notag
	\end{align}
	
	\item [(vi)]
	There is a strong connection between MIN based on Wigner-Yanase skew information and our measure, that is $\mathcal{N}^b_s(\rho_{BA}\otimes \rho_{AB}) \geq \mathcal{N}_s(\rho_{AB})$ since
	\begin{align}
		&\mathcal{N}^b_s(\rho_{BA}\otimes \rho_{AB}) \notag \\
		&=1-\mathop{min}\limits_{\{\Pi^{aa}_{st}\}}\sum\limits^m_{s=1}\sum\limits^m_{t=1}tr(\sqrt{\rho_{BA}\otimes \rho_{AB}}(\mathbb{I}_m\otimes\Pi^{aa}_{st}\otimes\mathbb{I}_v)\sqrt{\rho_{BA}\otimes \rho_{AB}}(\mathbb{I}_m\otimes\Pi^{aa}_{st}\otimes\mathbb{I}_v)\notag \\
		&\geq 1-\mathop{min}\limits_{\{\Pi^{a}_{s}\Pi^{a}_{t}\}}\sum\limits^m_{s=1}\sum\limits^m_{t=1}tr(\sqrt{\rho_{BA}\otimes \rho_{AB}}(\mathbb{I}_m\otimes\Pi^{a}_{s}\otimes\Pi^{a}_{t}\otimes\mathbb{I}_v)\sqrt{\rho_{BA}\otimes \rho_{AB}}(\mathbb{I}_m\otimes\Pi^{a}_{s}\otimes\Pi^{a}_{t}\otimes\mathbb{I}_v)\notag \\
		&=1-\mathop{min}\limits_{\{\Pi^{a}_{k}\}}(\sum\limits_k tr(\sqrt{\rho_{AB}}(\Pi^{a}_{k}\otimes\mathbb{I}_n)\rho_{AB}(\Pi^{a}_{k}\otimes\mathbb{I}_n))^2 \notag \\
		&\geq 1-\mathop{min}\limits_{\{\Pi^{a}_{k}\}}\sum\limits_k tr(\sqrt{\rho_{AB}}(\Pi^{a}_{k}\otimes\mathbb{I}_n)\rho_{AB}(\Pi^{a}_{k}\otimes\mathbb{I}_n)) \notag \\
		&=\mathcal{N}_s(\rho_{AB}).\notag
	\end{align}
\end{itemize}

The nonbilocality measure for input any pure states can evaluated as follows.

{\bf Theorem 1.}\label{thm1}
If $\rho_{AB}\otimes\rho_{CD}=\ket{\psi_{AB}}\bra{\psi_{AB}}\otimes\ket{\phi_{CD}}\bra{\phi_{CD}}$ is pure, that is $\ket{\psi_{AB}}=\sum_i\lambda_i\mid i_ai_b\rangle$ and $\ket{\phi_{CD}}=\sum_j\mu_j\mid j_cj_d\rangle$, then
	\begin{equation}
		\mathcal{N}^b_s(\rho_{AB}\otimes \rho_{CD})=1-\sum_{ij}\lambda_i^4\mu_j^4.
	\end{equation}

{\bf Proof}\label{p1}
First, noting that any von Neumann measurements on $H_B\otimes H_C$ is expressed as $\Pi^{BC}=\{\Pi^{BC}_{st}\equiv U\mid s_bt_c\rangle\langle s_bt_c\mid U^\dagger\}$ and thus
\begin{align}
	\rho_{AB}\otimes \rho_{CD}=\sum\limits_{ii^\prime jj^\prime}\lambda_i\lambda_{i^\prime}\mu_j\mu_{j^\prime} &\mid i_a\rangle\langle i^\prime_a\mid\otimes \mid i_b\rangle\langle i^\prime_b\mid\otimes \mid j_c\rangle\langle j^\prime_c\mid\otimes \mid j_d\rangle\langle j^\prime_d\mid\notag
\end{align}
	and
	$$
	\rho_{BC}=tr_{AD}(\rho_{AB}\otimes \rho_{CD})=\sum\limits_{ij}\lambda^2_i\mu^2_j\mid i_bj_c\rangle\langle i_bj_c\mid
	$$
	
	\begin{align}
		&\sqrt{\rho_{AB}\otimes \rho_{CD}}(\mathbb{I}_m\otimes\Pi^{BC}_{st}\otimes\mathbb{I}_v)\sqrt{\rho_{AB}\otimes \rho_{CD}}(\mathbb{I}_m\otimes\Pi^{BC}_{st}\otimes\mathbb{I}_v)\notag \\
		&=\sum\limits_{iji^\prime j^\prime}\sqrt{\lambda_i\lambda_{i^\prime}\mu_j\mu_{j^\prime}} \mid i_a\rangle\langle i^\prime_a\mid\otimes \mid i_b\rangle\langle i^\prime_b\mid \otimes  \mid j_c\rangle\langle j^\prime_c\mid\otimes \mid j_d\rangle\langle j^\prime_d\mid U\mid s_bt_c\rangle\notag \\
		&\langle s_bt_c\mid U^\dagger\{\sum\limits_{uvu^\prime v^\prime} \sqrt{\lambda_u\lambda_{u^\prime}\mu_v\mu_{v^\prime}} \mid u_a\rangle\langle u^\prime_a\mid \otimes\mid u_b\rangle\langle u^\prime_b\mid \otimes \mid v_c\rangle\langle v^\prime_c\mid\otimes \mid v_d\rangle\langle v^\prime_d\mid \} U\mid s_bt_c\rangle\langle s_bt_c\mid U^\dagger\notag \\
		&=\sum\limits_{iji^\prime j^\prime}\sum\limits_{uvu^\prime v^\prime} \sqrt{\lambda_i\lambda_{i^\prime}\mu_j\mu_{j^\prime} \lambda_u\lambda_{u^\prime}\mu_v\mu_{v^\prime}} \mid i_a\rangle\langle i^\prime_a\mid u_a\rangle\langle u^\prime_a\mid \otimes\mid i_bj_c\rangle\notag \\
		& \langle i^\prime_b j^\prime_c\mid U\mid s_bt_c\rangle\langle s_bt_c\mid U^\dagger \mid u_bv_c\rangle\langle u^\prime_b v^\prime_c\mid U\mid s_bt_c\rangle\langle s_bt_c\mid U^\dagger \otimes\mid j_d\rangle\langle j^\prime_d\mid v_d\rangle\langle v^\prime_d\mid\notag
	\end{align}

from which we have
	\begin{align}
		&\sum\limits_{sv}tr(\sqrt{\rho_{AB}\otimes \rho_{CD}}(\mathbb{I}_m\otimes\Pi^{BC}_{st}\otimes\mathbb{I}_v) \sqrt{\rho_{AB}\otimes \rho_{CD}}(\mathbb{I}_m\otimes\Pi^{BC}_{st}\otimes\mathbb{I}_v)\notag \\
		&=\sum\limits_{sv}\sum\limits_{iujv} \lambda_i \lambda_u \mu_j \mu_v \langle u_bv_c\mid U\mid s_bt_b\rangle\langle s_bt_b\mid U^\dagger \mid u_bv_c\rangle\langle i_bj_c\mid U\mid s_bt_c\rangle\langle s_bt_c\mid U^\dagger \mid i_bj_c\rangle \notag \\
		&=\sum\limits_{sv}(\sum\limits_{ij} \lambda^2_i\mu^2_j   \langle i_bj_c\mid U\mid s_bt_c\rangle\langle s_bt_c\mid U^\dagger\mid i_bj_c\rangle)^2\notag \\
		&=\sum\limits_{sv}(\langle s_bt_c\mid U^\dagger \rho_{BC} U\mid s_bt_c\rangle)^2    \notag
	\end{align}
	
Since $\Pi^{BC}_{st}= U\mid s_bt_c\rangle\langle s_bt_c\mid U^\dagger$ leaves $\rho_{BC}$ invariant, it follows that
	
	$$
	\rho_{BC}=\sum_{st} U\mid s_bt_c\rangle\langle s_bt_c\mid U^\dagger \rho_{BC} U\mid s_bt_c\rangle\langle s_bt_c\mid U^\dagger
	$$
	
or, equivalently,
	$$
	\rho_{BC}=\sum_{st} \langle s_bt_c\mid U^\dagger \rho_{BC} U\mid s_bt_c\rangle U\mid s_bt_c\rangle\langle s_bt_c\mid U^\dagger
	$$
	
is a spectral decomposition of $\rho_{BC}$.
	
Consequently,
	\begin{align}
		&\sum\limits_{sv}tr(\sqrt{\rho_{AB}\otimes \rho_{CD}}(\mathbb{I}_m\otimes\Pi^{BC}_{st}\otimes\mathbb{I}_v) \sqrt{\rho_{AB}\otimes \rho_{CD}}(\mathbb{I}_m\otimes\Pi^{BC}_{st}\otimes\mathbb{I}_v)\notag \\
		&=\sum\limits_{sv}(\langle s_bt_c\mid U^\dagger \rho_{BC} U\mid s_bt_c\rangle)^2\notag \\
		&=\sum\limits_{st}\lambda^4_s\mu^4_t \notag
	\end{align}
The desired result is obtained. The optimum is achieved by any von Neumann measurement leaving $\rho_{BC}$ invariant. 

For mixed input states, there are some kinds of different situations and we listed as following:

Suppose the Hilbert spaces $H_A, H_B, H_C$ and $H_D$ are of dimensions $dim\ H_A=m$, $dim\ H_B=n$, $dim\ H_C=u$ and $dim\ H_D=v$, respectively. Let $L(H_i)$ be the Hilbert space consisting of all linear operators on $H_i (i=A,B,C,D)$, with the Hilbert-Schmidt inner product $\langle x\mid y\rangle\equiv tr(x^\dagger y)$. Let $\{X_i: i=0,1,\dots,m^2-1\}, \{Y_j: j=0,1,\dots,n^2-1\}, \{Z_k: k=0,1,\dots,u^2-1\}, \{W_l: l=0,1,\dots,v^2-1\}$ be orthonormal Hermitian operator bases for $L(H_A), L(H_B), L(H_C)$ and $L(H_D)$, respectively, with $X_0=\mathbb{I}^A/\sqrt{m}, Y_0=\mathbb{I}^B/\sqrt{n}, Z_0=\mathbb{I}^C/\sqrt{u}$ and $W_0=\mathbb{I}^D/\sqrt{v}$. Then, general bipartite states $\rho_{AB}$ and $\rho_{CD}$ can always be represented as

\begin{equation}\label{6}
	\sqrt{\rho_{AB}}=\sum\limits_{ij}t^{ab}_{ij}X_i\otimes Y_j,\qquad \sqrt{\rho_{CD}}=\sum\limits_{kl}t^{cd}_{kl}Z_k\otimes W_l
\end{equation}

where $t^{ab}_{ij}\equiv tr(\sqrt{\rho_{AB}}(X_i\otimes Y_j))$ and $t^{cd}_{kl}\equiv tr(\sqrt{\rho_{CD}}(Z_k\otimes W_l))$. Let $T_{ab}=(t^{ab}_{ij}), T_{cd}=(t^{cd}_{kl})$, which may be regarded as some kind of correlation matrices for the state $\rho_{AB}$ and $\rho_{CD}$, respectively. Then, we have

\begin{equation}
	\begin{split}
		\sqrt{\rho_{AB}\otimes \rho_{CD}}&=\sum\limits_{i,j,k,l} t^{ab}_{ij}t^{cd}_{kl}X_i\otimes Y_j\otimes Z_k\otimes W_l, \\
		\sqrt{\rho_{BC,AD}}&=\sum\limits_{j,k}\sum\limits_{i,l}(t^{ab}_{ij}t^{cd}_{kl})_{jk,il}Y_j\otimes Z_k\otimes X_i\otimes W_l,
	\end{split}
\end{equation}

where the matrix $T_{bc,ad} = (t^{bc,ad}_{jk,il})=(t^{ab}_{ij}t^{cd}_{il})=T^t_{ab}\otimes T_{cd}$

note that $\mathcal{N}^b_s(\rho_{AB}\otimes \rho_{CD}) =\mathcal{N}_s(\rho_{BC,AD})$, since 

\begin{align}
	tr(\sqrt{\rho_{AB}\otimes \rho_{CD}}(\mathbb{I}_m\otimes\Pi^{BC}_{st}\otimes\mathbb{I}_v)\sqrt{\rho_{AB}\otimes \rho_{CD}}(\mathbb{I}_m\otimes\Pi^{BC}_{st}\otimes\mathbb{I}_v)=tr(\sqrt{\rho_{BC,AD}}(\mathbb{I}_m\otimes\Pi^{BC}_{st}\otimes\mathbb{I}_v)\sqrt{\rho_{BC,AD}}(\mathbb{I}_m\otimes\Pi^{BC}_{st}\otimes\mathbb{I}_v)
\end{align}

{\bf Theorem 2.}\label{thm2}
	\rm{For $\rho_{AB}$ and $\rho_{CD}$ represented as eq.\eqref{6}, we have}
	\begin{equation}
		\mathcal{N}^b_s(\rho_{AB}\otimes \rho_{CD})=1-\mathop{min}\limits_{F}FT_{bc,ad}T^t_{bc,ad}F^t\leq 1-\sum\limits_{o=1}^{nu}t_o,
	\end{equation}
where $F \equiv (f_{g(jk)})$ is an $nu \times n^2u^2$-dimensional matrix with $f_{g(jk)}=tr(\Pi^{BC}_gY_j\otimes Z_k), (g=0,1,\dots,nu-1; (jk)=ju^2+k,j=0,1,\dots,n^2-1, k=0,1,\dots,u^2-1)$, $T_{bc,ad}=(t^{ab}_{ij}t^{cd}_{kl})=T^t_{ab}\otimes T_{cd}$ is an $n^2u^2\times m^2v^2$-dimensional matrix, and $\{t_o:o=1,2,\dots,n^2u^2\}$ are the eigenvalues of the matrix $T_{bc,ad}T^t_{bc,ad}$ listed in the increasing order.

{\bf Proof.}
	\begin{align}
		&\sum\limits_{st}(\mathbb{I}_m\otimes\Pi^{BC}_{st}\otimes\mathbb{I}_v)\sqrt{\rho_{BC,AD}}(\mathbb{I}_m\otimes\Pi^{BC}_{st}\otimes\mathbb{I}_v)\notag \\
		&=\sum\limits_g\sum\limits_{ijkl} t^{ab}_{ij}t^{cd}_{kl}\Pi^{BC}_g(Y_j\otimes Z_k)\Pi^{BC}_g\otimes X_i\otimes W_l\notag \\
		&=\sum\limits_g\sum\limits_{ijkl} t^{ab}_{ij}t^{cd}_{kl} f_{g(jk)} \Pi^{BC}_g\otimes X_i\otimes W_l\notag \\
		&=\sum\limits_g\sum\limits_{ijj^\prime kk^\prime l} t^{ab}_{ij}t^{cd}_{kl} f_{g(jk)} f_{g(j^\prime k^\prime)}Y_{j^\prime}\otimes Z_{k^\prime}\otimes X_i \otimes W_l\notag \\
	\end{align}
and therefore
	\begin{align}
		&\sum\limits_{st}tr(\sqrt{\rho_{BC,AD}}(\mathbb{I}_m\otimes\Pi^{BC}_{st}\otimes\mathbb{I}_v)\sqrt{\rho_{BC,AD}}(\mathbb{I}_m\otimes\Pi^{BC}_{st}\otimes\mathbb{I}_v)\notag \\
		&=\sum\limits_g \sum\limits_{ijj^\prime kk^\prime l} t^{ab}_{ij}t^{cd}_{kl} f_{g(jk)} f_{g(j^\prime k^\prime)} t^{ab}_{ij^\prime}t^{cd}_{k^\prime l}\notag \\
		&=\sum\limits_g \sum\limits_{ijj^\prime kk^\prime l} f_{g(jk)} t^{bc,ad}_{jk,il} t^{bc,ad}_{j^\prime k^\prime, il} f_{g(j^\prime k^\prime)}\notag \\
		&=FT_{bc,ad}T^t_{bc,ad}F^t
	\end{align}
By definition, we have
	\begin{equation}
		\mathcal{N}^b_s(\rho_{AB}\otimes \rho_{CD})=1-\mathop{min}\limits_{F}FT_{bc,ad}T^t_{bc,ad}F^t\leq 1-\sum\limits_{o=1}^{nu}t_o,
	\end{equation}
where $\{t_o:o=1,2,\dots,n^2u^2\}$ are the eigenvalues of the matrix $T_{bc,ad}T^t_{bc,ad}$ listed in the increasing order.

In addition, with out lose of generality, if $\rho_b$ is nondegenerate, then $\Pi^{BC}=\{\Pi^B_s \otimes \Pi^C_t\}=\{\mid s_b\rangle\langle s_b\mid \otimes \Pi^C_t\}$ and we have the following theorem.

{\bf Theorem 3.}\label{thm3}
If $\rho_b$ is nondegenerate, then
	\begin{equation}
		\begin{aligned}
			\mathcal{N}^b_s(\rho_{AB}\otimes \rho_{CD})=1-trBT^t_{ab}T_{ab}B^t\times \mathop{min}\limits_{C}trCT_{cd}T^t_{cd}C^t\leq1-trBT^t_{ab}T_{ab}B^t\times(\sum\limits_{o^\prime=1}^ut^\prime_{o^\prime})
		\end{aligned}
	\end{equation}
where $B\equiv (b_{sj})$ is an $n\times n^2$-dimensional matrix with $b_{sj}\equiv tr\mid s_b\rangle\langle s_b\mid Y_j$, $C\equiv(c_{tk})$ is a $u\times u^2$-dimensional matrix with $c_{tk}\equiv tr\Pi^c_tZ_k$ and $\{t^\prime_{o^\prime}: o^\prime= 1,2,\dots,u^2\}$ are the eigenvalues of the matrix $T_{cd}T^t_{cd}$ listed in increasing order.
	\par
In particular, if u=2, then
	\begin{equation}
		\mathcal{N}^b_s(\rho_{AB}\otimes \rho_{CD})=1-trBT^t_{ab}T_{ab}B^t\times(\Vert \boldsymbol{r}_{cd} \Vert^2+r^\prime)
	\end{equation}
where $\boldsymbol{r}_{cd}=(t^{cd}_{00}, t^{cd}_{01},\dots,t^{cd}_{0(v^2-1))})$ and $r^\prime$ is the smallest eigenvalue of the $3\times3$-dimensional matrix $RR^t$ with $R=(t_{kl})_{k=1,2,3;l=0,1,\dots,v^2-1}$.

{\bf Proof}
If $\rho_b$ is nondegenerate, we have
	\begin{align}
		&\mathcal{N}^b_s(\rho_{AB}\otimes \rho_{CD})=1-\mathop{min}\limits_{\{\Pi^{C}_{st}\}}\sum\limits^n_{s=1}\sum\limits^u_{t=1}tr(\sqrt{\rho_{AB}\otimes \rho_{CD}}(\Pi^B\otimes \Pi^C)\sqrt{\rho_{AB}\otimes \rho_{CD}}(\Pi^B\otimes \Pi^C)\notag \\
		&=1-\mathop{min}\limits_{\{\Pi^{C}_{g}\}}\sum\limits_{g}tr(\sqrt{\rho_{AB}}(\Pi^B)\sqrt{\rho_{AB}}(\Pi^B_g)tr(\sqrt{\rho_{CD}}(\Pi^C_g)\sqrt{\rho_{CD}}(\Pi^C_g)\notag \\
		&=1-\sum\limits_{g}tr(\sqrt{\rho_{AB}}(\Pi^B)\sqrt{\rho_{AB}}(\Pi^B_g)\mathop{min}\limits_{\{\Pi^{C}_{g^\prime}\}}\sum\limits_{g^\prime}tr(\sqrt{\rho_{CD}}(\Pi^C_{g^\prime})\sqrt{\rho_{CD}}(\Pi^C_g)\notag \\
		&=1-trBT^t_{ab}T_{ab}B^t\times \mathop{min}\limits_C trCT_{cd}T^t_{cd}C^t\notag \\
		&\leq 1-trBT^t_{ab}T_{ab}B^t\times (\sum\limits_{o^\prime=1}^ut^\prime_{o^\prime})\notag
	\end{align}
	\par
If u=2, the completeness relation $\sum^1_{t=0}\Pi^C_t=\mathbb{I}^C$ implies that $c_{0k}=-c_{1k}(k=1,2,3)$. Let $\boldsymbol{c}\equiv \sqrt{2}(c_{01},c_{02},c_{03})$, then from $\sum^3_{k=0}c^2_{0k}=1$ and $c_{00}=c_{10}=\frac{1}{\sqrt{2}}$, we get $\Vert \boldsymbol{c}\Vert$=1.
	\par
Then we have
	$$
	\begin{gathered}
		C=(c_{tk})=\frac{1}{\sqrt{2}}\begin{pmatrix} 1  &  \boldsymbol{c} \\ 1  &  -\boldsymbol{c}\end{pmatrix}
	\end{gathered}
	$$
and
	$$
	\begin{gathered}
		T_{cd}=(t_{kl})=\begin{pmatrix} \boldsymbol{r}_{cd} \\ R \end{pmatrix}
	\end{gathered}
	$$
with a $v^2$-dimensinal row vextor $\boldsymbol{r}_{cd}=(t^{cd}_{00}, t^{cd}_{01},\dots,t^{cd}_{0(v^2-1))})$, and a $3\times v^2$-dimensional matrix $R=(t_{kl})_{k=1,2,3;l=0,1,\dots,v^2-1}$. So we have
	$$
	trCT_{cd}T^t_{cd}C^t=\Vert \boldsymbol{r}_{cd} \Vert^2+\boldsymbol{c}RR^t\boldsymbol{r}^t
	$$

{\bf Theorem 4.}\label{thm4}
If the marginal states $\rho_b$ and $\rho_c$ are both nondegenerate, we strightforward have
	\begin{equation}
		\mathcal{N}^b_s(\rho_{AB}\otimes \rho_{CD})=1-trBT^t_{ab}T_{ab}B^t\times trCT_{cd}T^t_{cd}C^t
	\end{equation}
where $C\equiv (c_{tk})$ is a $u\times u^2$-dimensional matrix with $c_{tk}\equiv tr\mid t_c\rangle\langle t_c\mid  Z_k$. 

\section{Examples}
In this part, we calculate MINBS for the pure states and mixed states, respectively.

{\bf Example 1.}
	For any Bell states, such as $\mid \Phi_{AB}\rangle\otimes \mid \Phi_{CD}\rangle=\frac{1}{\sqrt{2}}(\ket{00}+\ket{11})_{AB}\otimes \frac{1}{2}(\ket{00} +\ket{11})_{CD}$, we have
	$$
	\mathcal{N}^b_s(\mid \Phi_{AB}\rangle\langle \Phi_{AB}\mid\otimes \mid \Phi_{CD}\rangle\langle \Phi_{CD}\mid)=1-4\times(\frac{1}{\sqrt{2}})^4\times(\frac{1}{\sqrt{2}})^4=\frac{3}{4}.
	$$

{\bf Example 2.}
	Consider the separable states, for the classical seperable state $\rho_{AB}=\rho_{CD}=\rho=\frac{1}{2}\ket{0}\bra{0}\otimes\ket{0}\bra{0}+\frac{1}{2}\ket{1}\bra{1}\otimes\ket{1}\bra{1}$, we have
	\begin{align}
		\sqrt{\rho}&=\frac{1}{\sqrt{2}}\ket{0}\bra{0}\otimes\ket{0}\bra{0}+\frac{1}{\sqrt{2}}\ket{1}\bra{1}\otimes\ket{1}\bra{1}\notag \\
		&=\frac{1}{\sqrt{2}}\frac{\mathbb{I}}{\sqrt{2}}\otimes\frac{\mathbb{I}}{\sqrt{2}}+\frac{1}{\sqrt{2}}\frac{\sigma_3}{\sqrt{2}}\otimes\frac{\sigma_3}{\sqrt{2}}\notag
	\end{align}
	\begin{gather}
		T_{ab}=T_{cd}=\begin{pmatrix} \frac{1}{\sqrt{2}} & 0 & 0 & 0 \\  0 & 0 & 0 & 0 \\ 0 & 0 & 0 & 0 \\ 0 & 0 & 0 & \frac{1}{\sqrt{2}}	\end{pmatrix}\notag
	\end{gather}
	We choose one of the most optimal von Neumann measurements as 
	$$
	\Pi^{BC}=\{H^{\otimes2}\ket{00}\bra{00}H^{\otimes2}, H^{\otimes2}\ket{01}\bra{01}H^{\otimes2}, H^{\otimes2}\ket{10}\bra{10}H^{\otimes2}, H^{\otimes2}\ket{11}\bra{11}H^{\otimes2}\},
	$$
	where $H$ denotes the Hadamard gate matrix
	\begin{gather}
		H=\frac{1}{\sqrt{2}}\begin{pmatrix} 1 & 1 \\ 1 & -1	\end{pmatrix}\notag
	\end{gather}
Through a strigtforward calculation we have
	$$
	\mathcal{N}^b_s(\rho\otimes \rho)=1-\mathop{min}\limits_{F}FT_{bc,ad}T^t_{bc,ad}F^t=\frac{3}{4}.
	$$

{\bf Example 3.} We consider the Bell-diagonal state $\rho_{ab}=\lambda_1\ket{\psi^+}\bra{\psi^+}+\lambda_2\ket{\psi^-}\bra{\psi^-}+\lambda_3(\ket{\phi^+}\bra{\phi^+}+\lambda_4\ket{\phi^-}\bra{\phi^-}$, where$\ket{\psi^\pm}=\frac{1}{\sqrt{2}}(\ket{00}\pm\ket{11})$ and $\ket{\phi^\pm}=\frac{1}{\sqrt{2}}(\ket{01}\pm\ket{10})$, $\sum\limits^3_{i=0}\lambda_i=1, \lambda_i\geq 0$; then by the operator functions,

$$
\sqrt{\rho_{AB}}=\sqrt{\lambda_1}\ket{\psi^+}\bra{\psi^+}+\sqrt{\lambda_2}\ket{\psi^-}\bra{\psi^-}+\sqrt{\lambda_3}\ket{\phi^+}\bra{\phi^+}+\sqrt{\lambda_4}\ket{\phi^-}\bra{\phi^-}
$$

Next we express the $\sqrt{\rho_{AB}}$ in the standard operator base $\{\frac{\mathbb{I}}{\sqrt{2}},\frac{\sigma_i}{\sqrt{2}}:i=1,2,3\}$

$$
\sqrt{\rho_{AB}}=\frac{h_0}{2}\frac{\mathbb{I}^a}{\sqrt{2}}\otimes\frac{\mathbb{I}^b}{\sqrt{2}}+\frac{h_1}{2}\frac{\sigma_1}{\sqrt{2}}\otimes\frac{\sigma_1}{\sqrt{2}}+\frac{h_2}{2}\frac{\sigma_2}{\sqrt{2}}\otimes\frac{\sigma_2}{\sqrt{2}}+\frac{h_3}{2}\frac{\sigma_3}{\sqrt{2}}\otimes\frac{\sigma_3}{\sqrt{2}},
$$

where 
\begin{align}
&h_0=\sqrt{\lambda_1}+\sqrt{\lambda_2}+\sqrt{\lambda_3}+\sqrt{\lambda_4}, \quad \ h_1=\sqrt{\lambda_1}-\sqrt{\lambda_2}+\sqrt{\lambda_3}-\sqrt{\lambda_4},\notag \\
&h_2=-\sqrt{\lambda_1}+\sqrt{\lambda_2}+\sqrt{\lambda_3}-\sqrt{\lambda_4},\quad h_3=\sqrt{\lambda_1}+\sqrt{\lambda_2}-\sqrt{\lambda_3}-\sqrt{\lambda_4},\notag
\end{align}

Thus we have

$$
T_{ab}=\begin{pmatrix} \frac{h_0}{2} & 0 & 0 & 0 \\ 0 & \frac{h_1}{2} & 0& 0 \\ 0 & 0 & \frac{h_2}{2}& 0 \\ 0 & 0 & 0& \frac{h_3}{2} \\ \end{pmatrix}
$$

and
$$
T_{aa,bb}=T^t_{ba}\otimes T_{ab}=diag(\frac{h_0^2}{4} ,\frac{h_0h_1}{4},\frac{h_0h_2}{4},\frac{h_0h_3}{4},\frac{h_1h_0}{4},\frac{h_1^2}{4},\frac{h_1h_2}{4},\frac{h_1h_3}{4},\frac{h_2h_0}{4},\frac{h_2h_1}{4},\frac{h_2^2}{4},\frac{h_2h_3}{4},\frac{h_3h_0}{4},\frac{h_3h_1}{4},\frac{h_3h_2}{4},\frac{h_3^2}{4})
$$

According to the Theorem 2, and we choose the von Neumann measurement
$$
\Pi^{BC}=\{\ket{\psi^+}\bra{\psi^+},\ket{\psi^-}\bra{\psi^-},\ket{\phi^+}\bra{\phi^+},\ket{\phi^-}\bra{\phi^-}\}
$$

and we can get the matrix
\addtocounter{MaxMatrixCols}{10}
$$
F=\frac{1}{2}\begin{pmatrix} 1 & 0 & 0 & 0 & 0 & 1 & 0 & 0 & 0 & 0 & -1 & 0 & 0 & 0 & 0 & 1 \\ 1 & 0 & 0 & 0 & 0 & -1 & 0 & 0 & 0 & 0 & 1 & 0 & 0 & 0 & 0 & 1  \\ 1 & 0 & 0 & 0 & 0 & 1 & 0 & 0 & 0 & 0 & 1 & 0 & 0 & 0 & 0 & -1  \\ 1 & 0 & 0 & 0 & 0 & -1 & 0 & 0 & 0 & 0 & -1 & 0 & 0 & 0 & 0 & -1 \\ \end{pmatrix}
$$
and 

$$
tr(FT_{aa,bb}T^t_{aa,bb}F^t)=\frac{(h_0^4+h_1^4+h_2^4+h_3^4)}{16}
$$

So we get the following elegant result
$$
\mathcal{N}^b_s(\rho_{ba}\otimes \rho_{ab})=1-\frac{(h_0^4+h_1^4+h_2^4+h_3^4)}{16}.
$$

The maximum $\frac{3}{4}$ is taken in any one of the four Bell states just like illustrated in Example 1 and the minimum equals to zero when $\lambda_i=\frac{1}{4}$ that is a produnct state, this result coincidence with the property (i).

We point out that when $c_1=\lambda_1-\lambda_2+\lambda_3-\lambda_4, c_2=-\lambda_1+\lambda_2+\lambda_3-\lambda_4, c_3=\lambda_1+\lambda_2-\lambda_3-\lambda_4$ are all equals to a parameter $-v$, where $v\in[-\frac{1}{3},1]$ , then the state $\rho_{ab}$ reduces to the two-qubit Werner state $\rho_{ab}=v\ket{\phi^-}\bra{\phi^-}+\frac{1-v}{4}\mathbb{I}^{ab}$, thus can applied our results to this situation.

\section{Conclusions}
In this article, we have proposed a new form of measurement-induced nonbilocal correlation measure based on Wigner-Yanase skew information. Then we demonstrate there is a connection between MINS and our measure, and we have presented an analytical formulas of MINBS for pure input states and provide upper bounds for mixed input states. It's interesting for the example we have proposed and it's natural to ask whether there exist correlations can make our measure get one, ie., have the maximal nonbilocal correlation. An interesting future work is to explore a effective way to measure the measurement-induced non N-local correlation for N larger than two.

\section*{Acknowledgement}
This work is supported by the National Natural Science Foundation of China, (Grant Nos. 11575155,  12165020).

\end{document}